# Brain network efficiency is influenced by pathological source of corticobasal syndrome

John D. Medaglia[1], Weiyu Huang[2], Santiago Segarra[2],

Christopher Olm[3], James Gee[4], Murray Grossman[3], Alejandro Ribeiro[2],

Corey T. McMillan[3], Danielle S. Bassett*[2,5]

[1]Department of Psychology, University of Pennsylvania

Philadelphia, PA, 19104, USA

[2]Department of Electrical and Systems Engineering, University of Pennsylvania

Philadelphia, PA, 19104, USA

[3]Department of Neurology, University of Pennsylvania

Philadelphia, PA, 19104, USA

[4]Penn Image Computing and Science Lab, University of Pennsylvania

Philadelphia, PA, 19104, USA

[5]Department of Bioengineering, University of Pennsylvania

Philadelphia, PA, 19104, USA

* To whom correspondence should be addressed;

Danielle S. Bassett

240 Skirkanich Hall

University of Pennsylvania

Philadelphia, PA, 19104

E-mail: dsb@seas.upenn.edu

Running title: Local efficiency in corticobasal syndrome

## Abstract

Corticobasal syndrome is a neurodegenerative condition that can be caused by either frontotemporal lobar degeneration or Alzheimer's disease pathology. Multimodal neuroimaging studies using volumetric MRI and DTI successfully discriminate between Alzheimer's disease and frontotemporal lobar degeneration. However, this evidence has typically included clinically heterogeneous patient cohorts and has rarely assessed the network structure of these distinct sources of pathology. Here we assess the ability of neuroimaging to discriminate between pathological sources in a clinically homogeneous cohort of patients with corticobasal syndrome using graph theoretical methods sensitive to alterations in distributed brain circuits. Pathology was confirmed by autopsy or a pathologically validated cerebrospinal fluid total tau-to-beta-amyloid ratio (T-tau/A$\beta$). Using structural MRI data, we identify association areas in fronto-temporo-parietal cortex with reduced gray matter density in corticobasal syndrome (N = 40) relative to age-matched controls (N = 40). Using these fronto-temporo-parietal regions of interest, we construct structural brain networks in subgroups of individuals with frontotemporal lobar degeneration (N = 19) or Alzheimer's disease (N = 21) pathology by linking these regions by the number of white matter streamlines identified in a deterministic tractography analysis of diffusion tensor imaging data. We characterize these structural networks using 5 graph-based statistics, and assess their relative utility in classifying underlying pathology. To evaluate classification power, we apply leave-one-out cross validation using a supervised support vector machine for each network statistic separately, as well as for gray matter density. The support vector machine procedure demonstrates that gray matter density poorly discriminates between frontotemporal lobar degeneration and Alzheimer's disease pathology subgroups with low sensitivity (57%) and specificity (52%). In contrast, a statistic of local network efficiency demonstrates excellent discriminatory power, with 85% sensitivity and 84% specificity. Our results indicate that the underlying pathological sources of corticobasal syndrome can be classified more accurately using graph theoretical statistics derived from patterns of white matter microstructure in association cortex than by regional gray matter density alone. These results highlight the importance of a multimodal neuroimaging approach to diagnostic analyses of corticobasal syndrome and suggest that distinct sources of pathology mediate the circuitry of brain regions affected by corticobasal syndrome.

# Introduction

Network science offers methods to analyze complex relational data. Recently, these methods have been applied to neuroimaging data acquired in clinical populations, where they can complement more traditional neuroimaging analyses by offering a characterization of the effects of progressive neuropathology on brain organization – effects that may be indiscernible at the level of single voxels, regions, or white matter tracts. In the context of disease, striking insights resulting from the applications of these methods to neuroimaging data include the identification of circuit-level predictors of cognitive decline (Crossley et al., 2014, Stam, 2014, Warren et al., 2014). Such observations beg the question of how exactly the degradation of distinct brain networks might lead to overlapping clinical disorders. Recent evidence suggests that network hubs play critical roles in normative processes associated with cognitive and motor function (Buckner et al., 2009, Medaglia et al., 2015, van den Heuvel and Sporns, 2013), and that their failure leads to abnormal symptomatology (Bullmore and Sporns, 2012, Ibrahim et al., 2014, Seeley et al., 2009, Warren et al., 2014). Yet, specific neuropathological mechanisms linking network properties to observable behavioral phenotypes in clinical disorders have remained elusive.

A particularly salient clinical presentation resulting from progressive neuropathology is corticobasal syndrome. Corticobasal syndrome is a progressive neurodegenerative condition in which cognitive and motor decline can be construed as a gradual failure of structural brain networks and the dynamics they support. Clinically, corticobasal syndrome is characterized by a lateralized motor disorder, including features such as rigidity, tremor, myoclonus, and limb dystonia, as well as cortical sensory loss. In addition, cognitive symptoms may include slowed speech associated with apraxia of speech or agrammatism, visuospatial deficits such as lateralized neglect, imparied perceptual organization, and spatial difficulty, deficits in executive functioning such as cognitive inflexibility, limited working memory and impaired word fluency, and a disorder of social cognition and personality change (Armstrong et al., 2013). Corticobasal syndrome can be caused by two distinct neuropathologies: (i) a form of frontotemporal lobar degeneration associated with the accumulation of misfolded and hyperphosphorylated tau, or (ii) Alzheimer's disease with deposits of beta-amyloid in the form of neuritic plaques and paired helical filaments of tau that result in neurofibrillary tangles (see also (Armstrong et al., 2013, Whitwell et al., 2010)). In an era of disease-modifying treatments that target specific pathologic species, it is critical to define the specific histopathologic abnormality in an individual with corticobasal syndrome. This combination of differing pathophysiological mechanisms has led to considerable challenges in diagnosis and treatment. At present, the pathology of underlying corticobasal syndrome is predicted antemortem in only 25-56% of cases (Armstrong et al., 2013).

While clinical features rarely differ between distinct sources of pathology in corticobasal syndrome, large-scale neuroimaging data offer a potential key to predicting corticobasal syndrome pathology antemortem, thus informing clinical treatments and interventions. Indeed, recent studies have demonstrated that neuroimaging is useful in discriminating between frontotemporal lobar degeneration and Alzheimer's disease pathology in clinical cohorts that may include some patients with corticobasal syndrome. Specifically, regional analyses of both gray matter density and white matter tracts revealed reductions of anterior temporal cortex in frontotemporal lobar degeneration compared to Alzheimer's disease and in posterior cingulate and precuneus in Alzheimer's disease relative to frontotemporal lobar degeneration (McMillan et al., 2012, McMillan et al., 2014). Individuals with Alzheimer's disease or frontotemporal lobar degeneration pathology presenting as corticobasal syndrome demonstrate shared loss of premotor, supplementary motor, and insular gray matter density. In addition, these pathologic groups have been shown to present with widespread temporoparietal and fronto-temporo-parietal loss, respectively (Whitwell et al., 2010). Other work has shown significantly greater white matter change in individuals with frontotemporal lobar degeneration pathology compared to those with Alzheimer's disease pathology (McMillan et al., 2012). However, this prior evidence has been constrained in two ways. First, cohorts assessed in these studies were clinically heterogeneous, being constituted by several clinical syndromes including but not limited to corticobasal syndrome. Thus, specific neuroimaging biomarkers that directly predict frontotemporal lobar degeneration or Alzheimer's disease may have been confounded by the distribution of disease associated with these phenotypes. For example, atypical forms of Alzheimer's disease such as logopenic variant primary progressive aphasia have contributed to a more posterior distribution of GM disease contributing to an Alzheimer's pathological diagnosis, while frontal white matter disease often observed in nonfluent primary progressive aphasia associated with tau pathology may have contributed to the observed association of anterior white matter disease with a frontotemporal lobar degeneration pathological diagnosis. Second, to date multimodal discrimination between FTLD and AD pathology has focused on neuroanatomic (McMillan et al., 2012) or data-driven (McMillan et al., 2013, McMillan et al., 2014) regional measures and thus have rarely evaluated the circuitry and interaction between candidate network structures.

Here, we address these prior gaps by applying a network analytic technique to multimodal neuroimaging data to differentiate pathological drivers of corticobasal syndrome. We hypothesize that frontotemporal lobar degeneration and Alzheimer's disease pathology may contribute to subtle differences in the degeneration of fronto-temporo-parietal regions – key loci of multiple cognitive hubs implicated in corticobasal syndrome – and that this is discernible as differential network patterns. Specifically, we expect widespread gray matter density reductions in frontal, parietal and temporal lobes in patients with corticobasal syndrome (Whitwell et al., 2010). However, we predict that the sensitivity of gray matter analyses is limited in discriminating pathology in a sample with a clinically homogeneous syndrome. We hypothesize instead that the network characteristics associated with areas with reduced gray matter density in CBS display discriminable patterns in frontotemporal lobar degeneration versus Alzheimer's disease pathology, representing complex pathological consequences of these diseases. We test these predictions in a clinically homogeneous population with corticobasal syndrome using diffusion-weighted images in participants with frontotemporal lobar degeneration or Alzheimer's disease pathology confirmed by autopsy or autopsy-validated cerebrospinal fluid tau/beta-amyloid ratios. Using patterns of network statistics evaluated at individual brain regions, we train support vector machine classifiers to categorize frontotemporal lobar degeneration and Alzheimer's disease pathology, and we identify statistics with the highest discriminative power. Our results offer a predictive multimodal biomarker of the underlying pathological sources of corticobasal syndrome based on the structural network architecture of association cortex.

## Materials and methods

### Participants

The patient cohort included 40 individuals from the Penn Frontotemporal Degeneration Center and Cognitive Neurology Clinic at the University of Pennsylvania who were clinically diagnosed with corticobasal syndrome and 40 age and sex matched elderly controls. A board-certified neurologist with extensive expertise in neurodegenerative diseases diagnosed all patients using published criteria (Armstrong et al., 2013). Written informed consent was obtained from all patients using a protocol approved by the University of Pennsylvania's Institutional Review Board. Alzheimer's disease or frontotemporal lobar degeneration pathology was confirmed by autopsy or cerebrospinal fluid analysis (see "Cerebrospinal fluid analysis" below for additional details). All patients participated in a multimodal MRI scanning session that included (i) a high-resolution volumetric T1-weighted MRI scan, (ii) a diffusion-weighted imaging protocol, and (iii) a lumbar puncture or post-mortem neuropathological exam. Patient groups were comparable on education, disease duration, and overall disease severity measured with the Mini-Mental Status Exam (MMSE), and Clinical Dementia Rating (CDR) scale (all p>0.1). Patients were also matched on frequency of clinician evaluation for the presence of apraxia, cortically mediated sensory loss, myoclonus, dystonia, visuospatial impairments, executive dysfunction, naming difficulty, and effortful speech. On average frontotemporal lobar degeneration patients were slightly older than Alzheimer's disease patients (t(91)=52.77; p=0.08; see Table 1). Although not significantly different, we adopted a conservative approach and examined the effect of age on classifier performance in addition to other features that can be associated with misclassifications (see Supplement); we found no relationship with any extrinsic demographic or clinical variable examined.

**Table 1. Demographics and Clinical Features**

**Demographics and Clinical Ratings**

|  | Age | Gender | MMSE | CDR* |
|---|---|---|---|---|
| **AD** | 60.3 (7.9) | 7M, 12F | 18.8 (7.3) | 2.5 (5.3) |
| **FTLD** | 65.7 (10.6) | 7M,13F | 22.1 (7.7) | 1.5 (3.3) |
| **Controls** | 60.9 (6.6) | 20M,20F | - | - |

**Clinical Symptom Frequency**

|  | Asymmetric Rigidity | Apraxia | Cortical Sensory Loss | Myoclonus | Dystonia |
|---|---|---|---|---|---|
| **AD** | 0.5 | 0.9 | 0.6 | 0.6 | 0.4 |
| **FTLD** | 0.9 | 0.8 | 0.5 | 0.3 | 0.4 |

|  | Visospatial Impairment | Executive Dysfunction | Naming Difficulty | Effortful Speech | Early Amnesia |
|---|---|---|---|---|---|
| **AD** | 0.8 | 0.8 | 0.8 | 0.3 | 0.4 |
| **FTLD** | 0.7 | 0.8 | 0.6 | 0.4 | 0.1 |

*CDR reported as mean, interquartile range. All others are mean (Standard Deviation). MMSE = Mini-Mental Status Exam, CDR = Clinical Dementia Rating.

## Cerebrospinal fluid and autopsy analysis

Consistent with our previous work (McMillan et al., 2014), cerebrospinal fluid (CSF) analyses of total tau (T-tau) and beta amyloid (Aβ) were obtained and evaluated with either a sandwich ELISA 2 (INNOTEST, Innogenetics, Ghent, Belgium) or a LUMINEX xMAP platform (INNO-BIA AlzBio3, Innogenetics). A ratio of total tau to beta–amyloid (T-tau/Aβ) was produced across platforms with an autopsy-validated conversion factor that has been cross-validated in two independent series (Irwin et al., 2012). Specifically, it has been demonstrated that a T-tau/Aβ ratio above threshold (0.34) is 95.5% accurate across two autopsy series in discriminating between frontotemporal lobar degeneration pathology and Alzheimer's disease pathology (Irwin et al., 2012). Using this ratio, we identified 15 patients with Alzheimer's disease pathology and 14 patients with a CSF profile consistent with frontotemporal lobar degeneration (Irwin et al., 2012, Armstrong et al., 2013).

The Integrated Neurodegenerative Disease Database was queried for neuropathological diagnoses and pathogenic genetic mutations, identified using previously reported procedures (Toledo et al., 2013), and this revealed a subset of 4 individuals with autopsy-confirmed Alzheimer's disease pathology, 4 individuals with frontotemporal lobar degeneration (2 progressive supranuclear palsy, 1 corticobasal degeneration) and 1 frontotemporal lobar degeneration with TDP-43 inclusions) pathology, and 4 individuals with pathogenic mutation associated with frontotemporal lobar degeneration (3 GRN, 1 MAPT).

Together, combining cerebrospinal fluid analyses, autopsy, and genetics, we identified a total cohort of 19 individuals with corticobasal syndrome with Alzheimer's disease and 21 patients with frontotemporal lobar degeneration.

## Volumetric T1 MRI acquisition and preprocessing

From each participant, we acquired a structural T1-weighted MPRAGE MRI using a Siemens Trio 3.0T scanner with an 8-channel phased-array head coil with the following parameters: repetition time (TR) = 1,620 ms, echo time (TE) = 3 ms, slice thickness = 1.0 mm, flip angle = 15°, matrix = 192 × 256, and in-plane resolution = 0.9 mm × 0.9 mm. T1 MRI images were preprocessed and gray matter density was calculated for each of 119 regions (see "Label construction" below) using the Advanced Normalization Tools (ANTS) (Avants et al., 2008, Avants et al., 2011) CorticalThickness package (Tustison et al., 2014). which incorporates the highly accurate (Klein et al., 2010) Advanced Normalization Tools (ANTs). Briefly, we used N4 bias correction to minimize image inhomogeneity effects. We then performed brain extraction using a combination of template-based and segmentation strategies involving registration of a dilated template brain that can then be used to guide brain segmentation from each individual MRI volume. Atropos six-tissue class segmentation (cortex, deep grey, brainstem, cerebellum, white matter, and CSF/other) was performed using an optimized combination of prior knowledge from N4 bias-correction and template-based priors. Voxelwise grey matter (GM) density measures were calculated as the weighted probability of a voxel belonging to a given tissue class. A diffeomorphic and symmetric algorithm in ANTs was then used to warp each GM density map to a custom template comprised of 115 controls and 93 neurodegenerative disease patients (Parkinson's disease, Alzheimer's disease, amyotrophic lateral sclerosis, and frontotemporal lobar degeneration) who are demographically comparable to the imaging series in the current study.

## Label construction

To define a common region definition for gray matter and network analyses, we created a label set using a multi-atlas label fusion (MALF) algorithm (Wang and Yushkevich, 2013) in ANTs on a labeled subset of the OASIS data set (Neuromorphometrics, Inc. http://Neuromorphometrics.com/, (Klein and Tourville, 2012)) under academic subscription. The OASIS labels are based on expert labeling and independent sampling of individual subject brains. Labels not included across all subjects were excluded from the labeling procedure. We normalized the labeled OASIS subset to the standard local template used for T1 processing and then we used the MALF procedure to generate a single consensus label set, consisting of 119 regions after the removal of white matter labels.

## Gray matter density analysis

We began by testing the hypothesis that association cortex in frontal, parietal, and temporal lobes is affected in corticobasal syndrome. We focused on gray matter density – a traditional neuroimaging measure – suggested to be reduced in association cortex across the sample. Specifically, we performed paired t-tests for gray matter density values in all 119 regions between corticobasal syndrome and healthy samples, followed by Bonferroni correction (Bonferroni, 1936) to identify regions with significantly reduced gray matter in corticobasal syndrome. We focus subsequent analysis of white matter network differences to these regions (see "Gray Matter Analysis" below).

## DTI acquisition and preprocessing

Diffusion-weighted images were acquired with either a 30- or 12-directional acquisition sequence. The 30-directional sequence (N = 30) included a single-shot, spin-echo, diffusion-weighted echo planar imaging sequence (field of view = 245 mm, matrix size 128 × 128; number of slices = 57; voxel size = 2.2 mm isotropic; TR = 6,700 ms; TE = 85 ms; and fat saturation). In total, 31 volumes were acquired per subject, one without diffusion weighting (b = 0 s/mm2) and 30 with diffusion weighting (b = 1,000 s/mm2) along 30 noncollinear directions. The 12-directional sequence (N = 10) included a single-shot, spin-echo, diffusion-weighted echo planar imaging sequence (field of view = 245 mm, matrix size = 128 × 128; number of slices = 40; voxel size = 1.7x1.7x3.0 mm; TR = 6,500 ms; and TE = 99 ms; and fat saturation). In total, 13 volumes were acquired per subject, one without diffusion weighting (b = 0 s/mm2) and 12 with diffusion weighting (b = 1,000 s/mm2) along 12 noncollinear directions. An equal proportion of DTI data from each sequence was available per subject group (, p>0.36). We additionally tested whether DTI directions were related to misclassifications and observed that they were not related (see Supplement). Diffusion-weighted images were preprocessed using ANTs (Avants et al., 2008, Avants et al., 2011) software. Briefly, the unweighted (b=0) images are first extracted and averaged. All DW images (including the individual b=0 volumes) were then aligned to the average b=0 using ANTs (Avants et al., 2008, Avants et al., 2011) An affine transform was applied to capture eddy distortion in the DW images as well as motion. Diffusion tensors were computed using a weighted linear least squares algorithm in Camino (Salvador et al., 2005). The corrected average b=0 image was aligned to the subject's T1 image from the same scanning session, first rigidly to correct for motion, then using a deformable diffeomorphic transformation with mutual information to correct for inter-modality distortion. The diffusion to T1 warp was composed with the T1 to template warp (from the cortical thickness pipeline), producing a mapping from DWI space to the population T1 template in a single interpolation. Tensors were resampled into the template space using log-Euclidean interpolation (Arsigny et al., 2006) and reoriented to preserve the anatomical alignment of white matter tracts (Alexander et al., 2001).

**Network methods**

See Figure 1 for a schematic and description of methods used to construct structural networks and train classifiers. See the Supplement for mathematical definitions of network features.

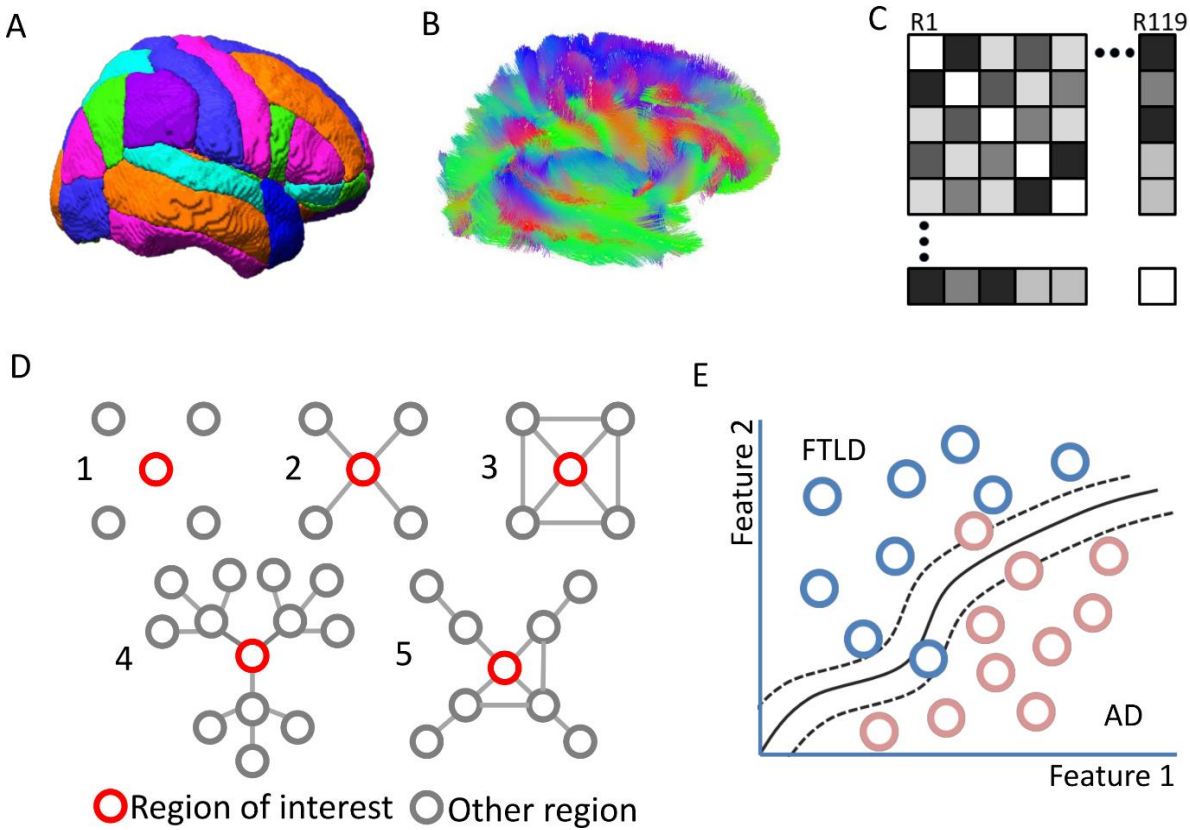

Figure 1: Schematic of the method. (A): Regions of interest (N = 119) were defined by OASIS labels registered to each individual's structural T1. (B): For each individual, we performed diffusion tractography to estimate streamlines connecting all voxel pairs. (C): An N X N adjacency matrix A whose elements Aij represent the number of streamlines reconstructed between region i and region j. We refer to each region as a network node, and each region-region connection as a network edge, weighted by the number of connecting streamlines. (D): We used 5 statistics at each node to classify frontotemporal lobar degeneration versus Alzheimer's disease: (1) gray matter density, which is agnostic to network connectivity, (2) node strength, which represents total edge weight of connections emanating from a region, (3) local clustering, which represents the extent to which a node's neighbors are also connected to each other, (4) eigenvector centrality, which is a statistic for the overall influence of a node in a network, and (5) local efficiency, which represents how connected the neighbors of a node are when this node is deleted. Note that in this illustration, the local efficiency for the node of interest is low. (E): Using the 5 statistics illustrated pictorially in panel (D), we apply a support vector machine to ntraining data (see "Support vector machine analysis") to determine the classification parameters and performance, equally weighted in sensitivity and specificity.

## DTI tractography

We performed diffusion tractography on each subject's diffusion tensor image using Camino (Cook et al., 2006). Fiber tracking began in each voxel of the label image and streamlines were generated using a linear fiber assignment by continuous tracking (FACT) (Mori and van Zijl, 2002). We terminated streamlines when either the fractional anisotropy in a voxel was less than 0.10, or the curvature of a streamline was greater than 75 degrees over a 5 mm span. We additionally used a set of labels consisting of regions that were slightly inflated to determine if allowing the catchment of additional streamlines aided in sensitivity to differences in underlying pathology (see Supplement).

## Streamline analysis

We examined which streamlines connecting pairs of regions were (1) reduced in corticobasal syndrome relative to elderly controls and (2) different between frontotemporal lobar degeneration and Alzheimer's disease. To do so, we performed paired t-tests for streamline counts in all region pairs with nonzero entries in at least one subject (N = 8,158 connections) between corticobasal syndrome and healthy samples, followed by Bonferroni correction (Bonferroni, 1936) to identify streamlines reduced in this sample.

## Network construction and regional statistics

Network science is a framework for representing and analyzing complex relational data (Newman, 2010). In this framework, components of a system are referred to as nodes, and connections between nodes are referred to as edges. Together, the nodes and the edges that connect them form a graph, which can be studied using techniques developed in the field of mathematics known as graph theory. To apply this network perspective to diffusion imaging data, we let a node represent a brain region, and we let an edge between two nodes represent the number of streamlines connecting the two brain regions. We do this for all brain regions and all connections between brain regions to construct a graph with N nodes and E edges.

A graph can be summarized in the form of an N×N adjacency matrix A. Here, we generated an adjacency matrix of size 119×119 for each participant. Each matrix element gives the number of streamlines connecting region i with region j. Using each individual's adjacency matrix, we calculated 5 commonly applied network statistics at each of the 119 brain regions. We selected the statistics based on the representation in the literature and theoretical relevance in their putative roles in mediating network dynamics. Specifically, we examined: (1) strength (sometimes referred to as "weighted degree", here defined as the sum of streamline counts to that particular region), (2) strength corrected for total edge weight in the network (also known as network density), (3) local clustering coefficient, (4) eigenvector centrality, and (5) local efficiency. See Supplement for mathematical definitions and (Rubinov and Sporns, 2010, Medaglia et al., 2015) for a discussion regarding the usage of network statistics in neuroimaging data.

Each statistic offers a unique quantification of the putative role that a brain region might play within the network. Strength represents the total number of weighted connections between a given node and all other nodes in the network. This statistic reflects how central a node is to the network without accounting for other features of network organization. Brain regions with high degree and strength are often considered to be "hubs" within the brain (Bullmore and Sporns, 2009) and are thought to serve crucial roles in mediating global network communication. By this definition, hub regions are relatively rare in brain networks: few hub regions exist alongside many sparsely connected regions. Strength divided by density adjusts these values by the total strength of connections in the network of each individual. We apply this density correction because average edge weight is agnostic to network topology but can directly affect classifier performance. We examine whether dividing strength by density improves classifier performance due to overall sampling of connections within the OASIS parcellation.

While strength and corrected strength are agnostic to local topology, the remaining 3 statistics directly assess local topology. For example, the local clustering coefficient represents the extent to which a node's neighbors are also connected to each other, therefore quantifying local node clustering within the network (Onnela et al., 2005). This represents the density of interactions between neighbors in the network. Eigenvector centrality is a statistic that reflects the influence of a node in a network, and is theoretically related to the region's role in the network's global dynamics (Lohmann et al., 2010). Intuitively, nodes with high eigenvector centrality are those that are connected to other important nodes – those that mediate communication between subcomponents of the network. Local efficiency represents how connected the neighbors of a node are when that node is deleted (Latora and Marchiori, 2001, Latora and Marchiori, 2003). This quantifies the robustness of local subnetworks to the removal of specific nodes, which can serve as a proxy for the importance of the node in local system failures (Latora and Marchiori, 2001, Latora and Marchiori, 2003) such as those observed in neurodegeneration.

## Support vector machine analysis

Support vector machine is a supervised learning method for binary classification (Suykens and Vandewalle, 1999), and is therefore often used to classify observations (e.g., patients) into two possible classes (e.g., frontotemporal lobar degeneration and Alzheimer's disease). We treat the following 3 types of data as features: gray matter density, white matter streamline counts connecting pairs of regions, and network statistics. We then train the support vector machines by providing them with labeled observations, for which the classification results are known. In traditional applications, an optimal linear classifier is built based on these labeled observations: the feature space is partitioned into two sectors (class 1 and class 2). Then, the performance of the classifier is tested by determining its accuracy on classifying a set of new observations – that is, observations that were not used during the training phase. However, the linear approach has significant disadvantages in the study of biological data, for which simple linear separations between diseased cohorts are uncommon. To address this limitation, we employ nonlinear classification using kernels (Rasmussen, 2006), which transform the feature space such that a linear classifier trained in the kernelized space is a non-linear classifier in the original feature space (see Figure 1E for an illustration of a non-linear classifier).

## Classifier training and testing

We evaluate classification power for gray matter density and network statistics calculated from cortical labels that displayed significantly reduced gray matter density in corticobasal syndrome in comparison to healthy controls. We assign a label (frontotemporal lobar degeneration or Alzheimer's disease) to each observation (subject in the corticobasal syndrome cohort) based on that subject's likely pathology using autopsy, genetic or CSF T-tau/Abeta ratio. We utilized labeled features for all observations except one to train a support vector machine classifier and compared the pathology predicted by the classifier for the left-out observation with his/her actual pathology (Arlot and Celisse, 2010). This process was repeated 40 times (i.e., the number of individuals in the clinical sample), each time with a different individual excluded from the training phase. We defined the global sensitivity and specificity of the classification to be the average performance over all trials for each feature.

The specific nonlinear classification approach that we used was built on radial basis kernels (Scholkopf et al., 1997), which include two free parameters: one for the kernel ($\gamma$) and one for the soft margin cost function (C). The $\gamma$ parameter quantifies the extent of the influence of each training observation in the construction of the classifier. Low values of $\gamma$ denote liberal or "far" influences, and high values denote conservative or "close" influences. The C parameter is a regularizing parameter preventing overfitting. It determines a trade-off between misclassification of the training examples and the simplicity of the decision surface. A low value of C results in a smooth decision surface that leads to a higher probability of misclassifications, whereas a high value of C results in a rougher decision surface that leads to a higher probability of overfitting. The impact of $\gamma$ and C on the performance of the classifier depends on the underlying features used in the classification. In the current context, we performed each leave-one-out cross validation over a range of $\gamma$ and C values ($\gamma$ from −15 to 3 in intervals of 0.5 and C from −5 to 15 in intervals of 0.5, totaling 1517 parameter pairs) (Chang and Lin, 2011). We defined the peak model performance as the maximum value shared by sensitivity and specificity in the $\gamma$ – C plane.

## Majority vote

Each network statistic can be sensitive to fine-scale differences in pathological drivers. Pragmatically speaking, we may wish to combine information from all statistics to maximize sensitivity to individual differences. We therefore constructed a combined classifier using a majority vote approach, which generally leads to better classification results by including more diverse information (Lam and Suen, 1997, Kuncheva and Whitaker, 2003). Specifically, each individual was assigned to either the predicted frontotemporal lobar degeneration group or the predicted Alzheimer's disease group according to the most frequently predicted assignment across the 5 network statistics.

## Results

### Gray matter density as a disease biomarker

To test the hypothesis that the gray matter in association areas in frontal, temporal, and parietal cortex would be more affected in corticobasal syndrome compared to healthy individuals, we tested for regional differences in gray matter density using two-tailed t-tests. We applied a Bonferroni correction (Bonferroni, 1936) for multiple comparisons and observed that 62 of the 119 regions displayed significantly less gray matter density in corticobasal syndrome in comparison to controls. These regions were anatomically located over a broad distribution of bilateral fronto-temporo-parietal cortex, including the primary and supplementary motor cortices, as well as the bilateral insula (see Figure 2 and Supplementary Table 1). We use these 62 areas as regions of interest in the following classification analysis.

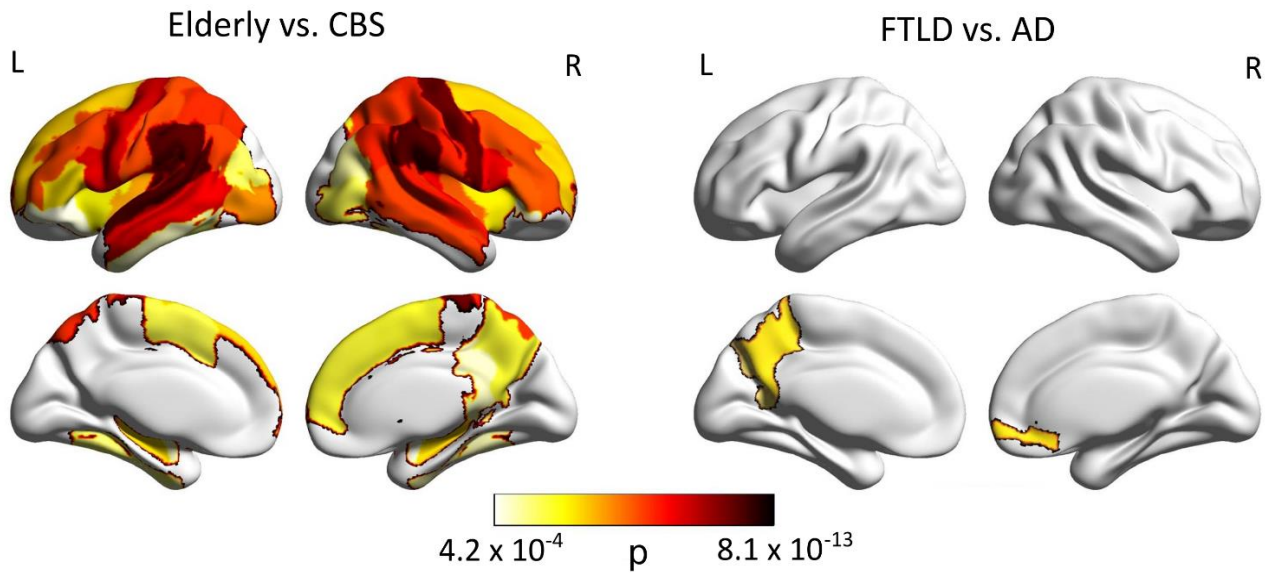

Figure 2: Gray matter differences between corticobasal syndrome & healthy controls, and between frontotemporal lobar degeneration (FTLD) & Alzheimer's disease (AD). Left Regions with significantly reduced gray matter density in individuals with corticobasal syndrome relative to controls following Bonferroni correction for multiple comparisons. A wide range of association regions within the frontal, parietal, and temporal lobes demonstrated reduced density in the diseased cohort. Right Regions with significantly different gray matter density in individuals with frontotemporal lobar degeneration relative to Alzheimer's disease. In all cases, volumes were reduced in Alzheimer's disease relative to frontotemporal lobar degeneration. Observe that only one region in the left precuneus and one region in the right medial frontal gyrus are found to be statistically significant following Bonferroni correction for multiple comparisons. In both panels, hotter colors indicate increasing statistical significance.

## White matter streamline differences as disease biomarkers

To contextualize our machine learning approach applied to network statistics, we tested for differences in individual streamlines connecting pairs of regions in (1) elderly individuals compared to those with corticobasal syndrome in addition to (2) individuals with frontotemporal lobar degeneration relative to controls. We applied a Bonferroni correction (Bonferroni, 1936) for multiple comparisons and observed that streamlines connecting 9 pairs of regions were reduced in corticobasal syndrome relative to controls predominantly within and between the right fronto-temporal cortex, in addition to one in the left frontal cortex and one interhemispheric connection between the left medial frontal cortex and right medial superior frontal gyrus (see Figure 3). There were no region pairs with significantly different streamline counts between individuals with frontotemporal lobar degeneration and Alzheimer's disease. See the Supplement for a list of all regions with reduced streamlines in corticobasal syndrome relative to controls.

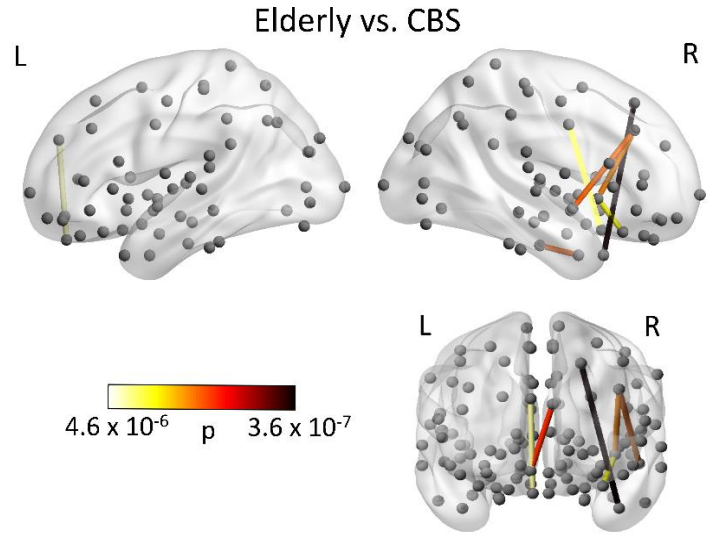

Figure 3: Streamline differences between corticobasal syndrome & healthy controls. Left Region pairs with significantly reduced white matter streamline counts in individuals with corticobasal syndrome relative to controls. Differences were most prominently observed in the right frontal and temporal cortices. Coronal representation is viewed facing the anterior surface. Hotter colors for connections indicate increasing statistical significance. See the Supplement for a list of regions with reduced streamlines.

## Network statistics as a pathology biomarker: a machine learning approach

We trained support vector machines on each of the 5 network statistics calculated for the 62 regions displaying reduced gray matter volume in corticobasal syndrome. We observed the best classification performance when using the local efficiency of regions as the features in the support vector machine: performance reached a peak sensitivity of 85% and a peak specificity of 84%. Other network statistics offered more modest sensitivities and specificities (See Figure 4).

To determine the relative utility of network statistics in comparison to univariate descriptors, we trained and tested support vector machines using either (i) gray matter density values for all regions, or (ii) all white matter streamlines connecting pairs of regions. The classifier based on regional gray matter density offered a maximum sensitivity of 57% and a maximum specificity of 52%. These results indicate that gray matter density measurements do not strongly distinguish between the two pathologies underlying corticobasal syndrome. We applied McNemar's test (McNemar, 1947) for correlated samples to examine differences in peak performance (maximum shared sensitivity/specificity) between network and non-network statistics. Only local efficiency demonstrated statistically superior performance to gray matter density ($\chi$) and to white matter streamline counts ($\chi$).

Three other network statistics demonstrated statistically significant classification performance relative to chance (50% classification) using a binomial sign test: strength (p = 0.008), strength corrected for network density (p = 0.003), and eigenvector centrality (p = 0.003).

The majority vote across network statistics offered equivalent performance to that obtained from the local efficiency alone (against gray matter: ($\chi$); against white matter: ($\chi$). Collectively, these results indicate that the 5 selected network statistics robustly classify frontotemporal lobar degeneration from Alzheimer's disease, but that local efficiency drives most of the classifier performance (see Figure 4; see "Relationship of network classification outcomes to extrinsic variables" in the Supplement for a discussion of misclassifications).

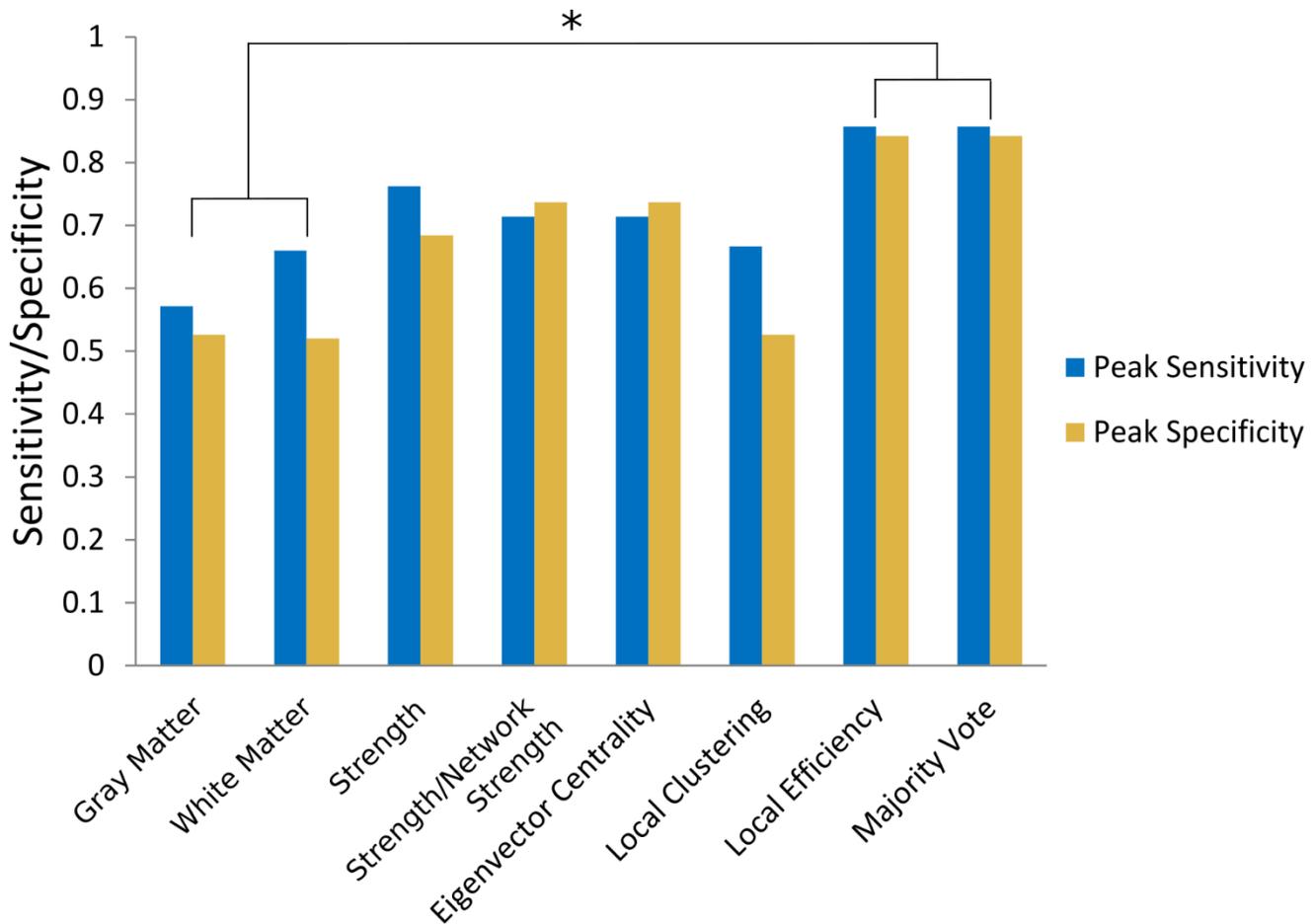

Figure 4: Discriminating power of different feature sets. Sensitivity (blue) and specificity (yellow) for gray matter density and white matter streamlines (left), network statistics (middle), and majority vote (right). Asterisk denotes that local efficiency and the majority vote perform significantly better than gray matter and white matter streamline classifiers. Observe that both sensitivity and specificity demonstrate approximately chance performance when only regional gray matter density is considered.

## Discussion

Our results demonstrate that gray matter imaging in conjunction with network techniques applied to white matter tractography can provide high accuracy in classifying the underlying neuropathology contributing to corticobasal syndrome. Specifically, we demonstrate that, within a clinically homogeneous group of patients with corticobasal syndrome, network architecture distinguishes between Alzheimer's disease and frontotemporal lobar degeneration pathology pathology in structural networks spanning a broad set of association regions in frontal, temporal, and parietal cortex. Our results indicate that structural network features can offer unique diagnostic value relative to other traditional neuroimaging measures in a sample where classification via clinical characteristics or neuroimaging is particularly challenging. Importantly, the higher-level information summarized by network statistics identifies meaningful variation in brain structure that cannot be observed with regional gray matter density measurements or streamline counts between pairs of regions. This is captured most strongly by "local efficiency," or the robustness of a local network to the degradation of a gray matter node. These findings serve as a basis for the development of multi-modal and multi-feature methods to quantify and predict the progression of heterogeneous disease processes in corticobasal syndrome. Moreover, they demonstrate that multimodal imaging, network science, and nonlinear machine learning can be used to quantify distinct pathologies underlying clinically homogeneous samples (McMillan et al., 2014).

Our novel findings are supported by an initial replication of the broad fronto-temporo-parietal gray matter loss in corticobasal syndrome due to Alzheimer's disease or other underlying pathology (Whitwell et al., 2010). In direct comparison of the two pathology subgroups, we demonstrate that gray matter density is reduced in the precuneus in Alzheimer's disease compared to frontotemporal lobar degeneration (McMillan et al., 2012) in a clinically homogeneous sample of individuals with corticobasal syndrome. Other work has shown significant gray matter disease in the precuneus in Alzheimer's disease relative to healthy controls (Acosta-Cabronero et al., 2010). This brain area is often associated with visual attention, a clinical characteristic of corticobasal syndrome. We also found significant gray matter atrophy in a ventral medial frontal distribution in Alzheimer's disease relative to tauopathy. While this medial frontal area is often noted as the focus of disease in patients with behavioral variant of frontotemporal degeneration (Schroeter et al., 2008), and while some patients with pathologically confirmed corticobasal degeneration may have disease in this region (Lee et al., 2011), this area is associated with a disorder of personality and social cognition, and the patients participating in this study did not have prominent changes in personality and cognition. Regardless of the clinical consequences of disease in these regions, the analysis of patterns of gray matter change in frontal, temporal, and parietal regions affected by corticobasal syndrome was not able to classify the corticobasal patients participating in this study on the basis of their underlying pathology. These results indicate that pathological processes affecting gray matter density are most discriminating between Alzheimer's disease and frontotemporal degeneration pathology in the right medial frontal gyrus and left precuneus, but gross patterns of gray matter density measurements in corticobasal syndrome are not sensitive and specific predictors of underlying pathology.

Within regions displaying reduced gray matter density in corticobasal syndrome, we examined the classification power of five network statistics extracted the associated white matter tractography data and found that local efficiency classified Alzheimer's disease versus frontotemporal lobar degeneration with 85% sensitivity and 84% specificity. The classification power of local efficiency was superior to that obtained from white matter streamlines connecting pairs of regions or from regional gray matter density. The performance of local efficiency was similar to a majority vote including the prediction of all five network statistics, suggesting that other characteristics of white matter network topology did not substantially contribute critical information to the use of MRI for the purpose of establishing the pathologic basis for corticobasal syndrome.

While other network features – strength, strength corrected for density, and eigenvector centrality – did not demonstrate superior performance to gray matter and pairwise white matter streamline counts in classifiers, they did demonstrate statistically significant classification value relative to chance. Local clustering demonstrated poorer performance overall; it is possible that this statistic can be affected by different forms of pathogenesis such as white matter specific pathological inclusions (McMillan et al., 2013) or non-specific processes such as Wallerian degeneration that can occur in both frontotemporal lobar degeneration and Alzheimer's disease.

**Implications for disease detection and classification**

Multimodal neuroimaging screening tools for Alzheimer's disease generally outperform cognitive measures (Ard, 2011) and can further reduce costs and increase efficiency for clinical trial entry by improving diagnostic accuracy (McMillan et al., 2014). Yet, understanding which features of imaging data offer promise as biomarkers of underlying neuropathologies has proven challenging. Techniques from graph theory and network science offer a potential solution to this challenge, by quantifying novel and complementary characteristics of brain organization (Bassett et al., 2011, Medaglia et al., 2015, Sporns, 2014). Indeed, network-based neuroimaging approaches are of increasing interest (Stam, 2014), not only for their potential in classifying neurological disorders and psychiatric disease (Finn et al., 2015) but also for their ability to reveal novel neurophysiological phenotypes and mechanisms. For example, for clinical identification, the cerebrospinal T-tau/A$\beta$ ratio can be used to discriminate Alzheimer's disease and frontotemporal lobar degeneration in corticobasal syndrome in vivo (Irwin et al., 2012). However, this ratio cannot enlighten us as to the complex structural anomalies underlying corticobasal syndrome and their potential implications for resilience and disease trajectory. With increasing emphasis on understanding the biological basis of cognitive dysfunction (of Mental Health, 2015), approaches involving a network perspective may be crucial to enhancing our understanding of the relationship between neuropathology and mental processes (Medaglia et al., 2015, Stam, 2014).

Our findings support prior theories positing that network-based statistics of neuroimaging data can provide unique diagnostic value in neurological samples (Anticevic et al., 2012, Bassett et al., 2008, Bassett et al., 2009, Bassett et al., 2012, Crossley et al., 2014, Ewers et al., 2011, Farb et al., 2013, Fillippi et al., 2013, Lynall et al., 2010, Stam, 2014, Wee et al., 2012). Our results demonstrate that an integrated analysis of structural network degradation can identify biomarkers for underlying pathologies that complement other neuroimaging measures. In particular, distinct patterns are observed in the local efficiency of association areas in frontal, temporal, and parietal cortex, which together share the majority of the pathological burdens associated with these diseases. These results demonstrate that certain network features may be more sensitive than others to underlying pathological processes. In other words, pathological burdens to the brain that appear superficially similar in gross neuroimaging approaches may nonetheless have dissociable effects on network topology. It is particularly informative to study patients with corticobasal syndrome because of the differences in white matter disease in the two pathologies implicated in corticobasal syndrome (McMillan et al., 2012). In Alzheimer's disease, white matter pathology is primarily due to Wallerian degeneration that follows from gray matter disease. Wallerian degeneration is also evident in frontotemporal lobar degeneration spectrum pathology. Moreover, the form of frontotemporal lobar degeneration associated specifically with tau pathology, the other major cause of corticobasal syndrome, is uniquely associated with specific white matter disease such as astrocytic plaques. This may play a role in local density, where compensation for limited white matter connectivity may be constrained by this unique source of white matter pathology. This form of white matter pathology may have less impact on other network statistics such as nodal strength, local clustering and eigenvector centrality that may be sensitive to Wallerian degeneration that is equally evident in both Alzheimer's disease and frontotemporal lobar degeneration pathology.

**Implications for cognitive resilience**

Our results bear additional relevance to emerging interests in the use of network techniques to understand cognitive function and dysfunction (Medaglia et al., 2015, Warren et al., 2014). In addition to providing information about mechanistic links between histopathological processes and clinical syndromes, network approaches can additionally enlighten us as to the nature of complex network failures underlying cognitive dysfunction (Warren et al., 2014, Stam, 2014). Recently, network techniques have shown promise in detecting cognitive trajectories that differentiate clinical presentations, such as in the conversion from mild cognitive impairment to Alzheimer's disease (Nir et al., 2015) using network measures of network so-called "small-worldness." However, the application of multimodal approaches and the examination of patterns of node-level network effects using machine learning have remained relatively limited (Richiardi et al., 2013).

Network degeneration in specific brain systems may differentially drive cognitive deficits. Indeed, several studies have found that association regions in frontal, temporal, and parietal cortices are replete with cognitive "hubs" that support a broad set of functions (Buckner et al., 2009, Medaglia et al., 2015, van den Heuvel and Sporns, 2013). Shifts in cognitive tasks are associated with modulations of the functional connectivity between these hub regions and other brain systems (Cole et al., 2013). Short path lengths, a marker of global network efficiency, in fronto-parietal systems are associated with increased general intelligence (van den Heuvel et al., 2009). These regions have also been implicated in the control of global brain dynamics (Gu et al., 2015), suggesting that they serve a crucial role in guiding the brain's overall performance in controlling cognition and behavior. Crucially, in neurological samples, failures in systems containing hubs is especially deleterious for global cognitive function (Warren et al., 2014, Stam, 2014). Thus, differences in patterns of progressive network neurodegeneration in corticobasal syndrome may impact cognitive resilience.

Here, we identify patterns of local efficiency in systems containing hubs that can be used to identify underlying pathology in a homogeneous clinical syndrome. Local efficiency is particularly interesting when considering cognitive resilience. In abstract physical models, this statistic measures the local tolerance of the network to a node's removal (Latora and Marchiori, 2001), and is thought to describe the importance of a node in information transfer in subnetworks. In other words, a subnetwork's vulnerability is related to the local efficiency of its contributing regions. If a subnetwork involves many nodes with low local efficiency, relatively minor perturbations will have drastic consequences for subnetwork function. This could bear implications for the differential consequences of Alzheimer's disease and frontotemporal degeneration pathology. If distinct tau pathology in subjects with frontotemporal degeneration is responsible for differences in network local efficiency, accumulating tauopathy in frontotemporal degeneration may eventually degrade network communities sufficiently to result in dissociable clinical disorders.

In particular, as white matter tracts degenerate, different subnetworks may fail in the two populations as a function of the interaction between accruing regional histopathology and underlying network local efficiency. As these distinct network failures evolve over time, clinical differences may become more salient and produce detectable cognitive and behavioral effects. In this context, it will be crucial in future to develop approaches that quantify mesoscale network organization in conjunction with neuropsychological measurements to track differentiating cognitive trajectories in corticobasal syndrome.

## Future directions

Our findings bear several implications for future work. As the field of network neuroscience matures, it may hold promise in the quantification of pathophysiological trajectories of neural systems in neurological diseases (Stam, 2014). In corticobasal syndrome, longitudinal follow-up studies could quantify progressive degeneration in Alzheimer's disease and frontotemporal lobar degeneration and determine to what extent their macro-scale network profiles diverge following initial presentation with corticobasal syndrome. In other neurological samples, evidence suggests that network structure (Raj et al., 2012) and function (Zhou et al., 2012) guides the spread of neurodegenerative processes. In particular, patterns of network vulnerability quantified by statistics such as local efficiency may predict pathology-driven degeneration later in the disease progression (Schmidt et al., 2016) that leads to distinct clinical presentations (Armstrong et al., 2013). In this regard, validating imaging based biomarkers may provide an important contribution to determining prognosis. It is interesting to consider that specific patterns of network pathology may provide unique information about clinical trajectory and probable efficacy of interventions.

In addition, the synthesis of neuroimaging measurements investigated here may provide a basis for multimodal and multiscale neuroimaging characterization of network degeneration in other behavioral variants that result from corticobasal degeneration (McMillan et al., 2014). This could provide an important contribution to ongoing developments in the conceptualization of corticobasal syndrome (Armstrong et al., 2013) in the nosology of clinical syndromes resulting from corticobasal degeneration. In particular, it may be possible to quantify sensitive and specific profiles of network neurodegeneration associated with frontal behavioral-spatial syndrome, nonfluent/agrammatic variant of primary progressive aphasia, and progressive supranuclear palsy syndrome (Armstrong et al., 2013).

## Methodological Considerations

It is important to mention a few methodological considerations pertinent to this work. First, only a subset of commonly applied network statistics was examined; other statistics may offer additional insights (Honey et al., 2010, Stam, 2014). In particular, network statistics that consider interdependencies of node roles (e.g., assortativity (Newman, 2010)) and describe features of the entire network organization (e.g., small world propensity (Muldoon et al., 2015)) may additionally provide sensitive and specific information. While we focused on commonly used node-level measures with distinct theoretical emphases, broadening analyses to consider the full breadth of available and novel statistics could prove informative.

Second, we used a parcellation of the brain into 119 (OASIS) regions using an atlas that has unique advantages in terms of expert labeling of individual brain scans applied with a validated consensus algorithm (Wang and Yushkevich, 2013). Other atlases with alternative approaches to labeling, including a priori and purely data-driven approaches (Evans et al., 2012), are expected to provide similar qualitative features (Bassett et al., 2011, de Reus and van den Heuvel, 2013, Wang et al., 2009) but may differ in quantitative statistics (Bassett et al., 2011). Third, we employed deterministic tractography to diffusion imaging data, and defined network edges as the number of streamlines connecting two regions (Hagmann et al., 2007, Bassett et al., 2011); other algorithms and probabilistic approaches may emphasize different features of white matter network organization in corticobasal syndrome (Dell'Acqua and Catani, 2012, Descoteaux et al., 2000). Fourth, classification accuracy may be enhanced in diffusion imaging techniques with greater resolution of diffusion directions. Finally, classification in the current study was in reference to the standard provided by tau/A$\beta$ ratios with autopsy confirmed pathology in part of the sample. However, there is the possibility that a small proportion of participants were incorrectly identified as Alzheimer's disease or frontotemporal lobar degeneration prior to the WM network based classification: the results of cross-validation suggest that about 5% of individuals may be incorrectly classified with the cerebrospinal fluid heuristic used here (Irwin et al., 2012).

## Conclusion

Local network topology in a distributed fronto-temporo-parietal system dissociates Alzheimer's disease from frontotemporal lobar degeneration in corticobasal syndrome. These results demonstrate the utility of multimodal network-based techniques in a clinically homogeneous cohort associated with corticobasal syndrome where there are two distinct underlying pathologies. Future research into patterns of WM network degeneration in corticobasal syndrome may help refine the nosology of neurodegeneration by providing a mechanistic link between histopathology and clinical syndromes mediated through brain network failure. Such a link could have extensive effects on diagnosis, the prediction of cognitive vulnerability and decline, and treatment planning in corticobasal syndrome.

# References

Acosta-Cabronero J, Williams GB, Pengas G, Nestor PJ. Absolute diffusivities define the landscape of white matter degeneration in alzheimer's disease. Brain 2010; 133: 529–539.

Alexander DC, Pierpaoli C, Basser PJ, Gee JC. Spatial transformations of diffusion tensor magnetic resonance images. IEEE Trans Med Imaging 2001; 20: 1131–1139.

Anticevic A, Cole MW, Murray JD, Corlett PR, Wang XJ, Krystal JH. The role of default network deactivation in cognition and disease. Trends Cogn Sci 2012; 16: 584–592.

Ard M C Edland SD. Power calculations for clinical trials in alzheimer's disease. J Alz Disease 2011; 26: 369–377.

Arlot S, Celisse A. A survey of cross-validation procedures for model selection. Stat Surveys 2010; 4: 40–79.

Armstrong MJ, Litvan I, Lang AE, Bak TH, Bhatia KP, Borroni B, et al. Criteria for the diagnosis of corticobasal degeneration. Neurology 2013; 80: 496–503.

Arsigny V, Fillard P, Pennec X, Ayache N. Log-euclidean metrics for fast and simple calculus on diffusion tensors. Magn Res Med 2006; 56: 411–421.

Avants BB, Epstein CL, Grossman M, Gee JC. Symmetric diffeomorphic image registration with cross-correlation: Evaluating automated labeling of elderly and neurodegenerative brain. Med Imag Anal 2008; 12: 26–41.

Avants BB, Tustison NJ, Song G, Cook PA, Klein A, Gee JC. A reproducible evaluation of ants similarity metric performance in brain image registration. NeuroImage 2011; 54: 2033–2044.

Bassett DS, Bullmore ET, Meyer-Lindenberg A, Apud JA, Weinberger DR, Coppola R. Cognitive fitness of cost-efficient brain functional networks. Proc Natl Acad Sci 2009; 106: 11747–11752.

Bassett DS, Bullmore ET, Verchinski BA, Mattay VS, Weinberger DR, Meyer-Lindenberg A. Hierarchical organization of human cortical networks in health and schizophrenia. J Neurosci 2008; 28: 9239–9248.

Bassett DS, Nelson BG, Mueller BA, Camchong J, Lim KO. Altered resting state complexity in schizophrenia. NeuroImage 2012; 59: 2196–2207.

Bassett DS, Wymbs NF, Porter MA, Mucha PJ, Carlson JM, Grafton ST. Dynamic reconfiguration of human brain networks during learning. Proc Natl Acad Sci 2011; 108: 7641–7646.

Bonferroni CE. Teoria statistica delle classi e calcolo delle probabilita. Pub R Istitut Sup Sci Econ Commerc Fir 1936; 8: 1–62.

Buckner RL, Sepulcre J, Talukdar T, Krienen FM, Liu H, Hedden T, et al. Cortical hubs revealed by intrinsic functional connectivity: mapping, assessment of stability, and relation to Alzheimer's disease. J Neurosci 2009; 29: 1860–1873.

Bullmore E, Sporns O. Complex brain networks: Graph theoretical analysis of structural and functional systems. Nat Rev Neurosci 2009; 10: 186–198.

Bullmore E, Sporns O. The economy of brain network organization. Nat Rev Neurosci 2012; 13: 336–349.

Chang CC, Lin CJ. Libsvm: A library for support vector machines. ACM T Intel Sys Tech 2011; 2.

Cole MW, Reynolds JR, Power JD, Repovs G, Anticevic A, Braver TS. Multi-task connectivity reveals flexible hubs for adaptive task control. Nat Neurosci 2013; 16: 1348–1355.

Cook P, Bai Y, Nedjati-Gilani S. Camino: Open-source diffusion-mri reconstruction and processing. Presented at the 14th Scientific Meeting of the International Society for Magnetic Resonance Imaging in Medicine 2006.

Crossley NA, Mechelli A, Scott J, Carletti F, Fox PT, McGuire P, et al. The hubs of the human connectome are generally implicated in the anatomy of brain disorders. Brain 2014; 137: 2382–2395.

de Reus MA, van den Heuvel MP. The parcellation-based connectome: Limitations and extensions. Neuroimage 2013; 80: 397–404.

Dell'Acqua F, Catani M. Structural human brain networks: hot topics in diffusion tractography. Curr Opin Neurol 2012; 25: 375–383.

Descoteaux M, Deriche R, Knosche TR, Anwander A. Deterministic and probabalistic tractography based on complex fibre orientation distributions. IEEE T Med Imag 2000; 28: 286.

Evans AC, Janke AL, Collins DL, Baillet S. Brain templates and atlases. NeuroImage 2012; 62: 911–922.

Ewers M, Sperling RA, Klunk WE, Weiner MW, Hampel H. Neuroimaging markers for the prediction and early diagnosis of alzheimer's disease dementia. Trends Cogn Sci 2011; 34: 430–442.

Farb NAS, Grady CL, Strother S, Tang-Wai DF, Masellis M, Black S, et al. Abnormal network connectivity in frontotemporal dementia: Evidence for prefrontal isolation. Cortex 2013; 49: 1856–1873.

Fillippi M, Agosta F, Scola E, Canu E, Magnani G, Marcone A, et al. Functional network connectivity in the behavioral variant of frontotemporal dementia. Cortex 2013; 49: 2389–2401.

Finn ES, Shen X, Scheinost D, Rosenberg MD, Huang J, Chun MM, et al. Functional connectome fingerprinting: identifying individuals using patterns of brain connectivity. Nat Neurosci 2015; 11: 1664.

Gu S, Pasqualetti F, Cieslak M, Telesford QK, Yu AB, Kahn AE, et al. Controllability of structural brain networks. Nat Comm 2015; 6: 8414.

Hagmann P, Kurant M, Gigandet X, Thiran P, Wedeen V, Meuli R. Mapping human whole-brain structural networks with diffusion mri. PLoS ONE 2007; 2: e597.

Honey CJ, Thivierge JP, Sporns O. Can structure predict function in the human brain? NeuroImage 2010; 52: 766–776.

Ibrahim GM, Cassel D, Morgan BR, Smith ML, Otsubo H, Ochi A, et al. Resilience of developing brain networks to interictal epileptiform discharges is associated with cognitive outcome. Brain 2014; 137: 2690–2702.

Irwin DJ, McMillan CT, Toledo JB, Arnold SE, Shaw LM, Wang LS, et al. Comparison of cerebrospinal fluid levels of tau and ab in alzheimer disease and frontotemporal degeneration using 2 analytical platforms. Arch Neurol 2012; 69: 1018–1025.

Klein A, Ghosh SS, Avants B, Yeo BTT, Fischl B, Ardekani B, et al. Evaluation of volume-based and surface-based brain image registration methods. NeuroImage 2010; 51: 214–220.

Klein A, Tourville J. 101 labeled brain images and a consistent human cortical labeling protocol. Front Neurosci 2012; 6: 171.

Kuncheva LI, Whitaker CJ. Measures of diversity in classifier ensembles and their relationship with the ensemble accuracy. Mach Learn 2003; 51: 181–207.

Lam L, Suen CY. Application of majority voting to pattern recognition: An analysis of its behavior and performance. IEEE T Sys Man Cyb 1997; 27: 468–480.

Latora V, Marchiori M. Efficient behavior of small-world networks. Phys Rev Lett 2001; 87: 198701.

Latora V, Marchiori M. Economic small-world behavior in weighted networks. Eur Phys J B 2003; 32: 249–263.

Lee SE, Rabinovici GD, Mayo MC, Wilson SM, Seeley WW, DeArmond SJ, et al. Clinicopathological correlations in corticobasal degeneration. Ann Neurol 2011; 70: 327–340.

Lohmann G, Margulies DS, Horstmann A, Pleger B, Lepsien J, Goldhahn D, et al. Eigenvector centrality mapping for analyzing connectivity patterns in fmri data of the human brain. PLoS One 2010; 5: e10232.

Lynall ME, Bassett DS, Kerwin R, McKenna P, Muller U, Bullmore ET. Functional connectivity and brain networks in schizophrenia. J Neurosci 2010; 30: 9477–9487.

McMillan CT, Avants B, Irwin DJ, Toledo JB, Wolk DA, Van Deerlin VM, et al. Can mri screen for csf biomarkers in neurodegenerative disease? Neurology 2013; 80: 132–138.

McMillan CT, Avants BB, Cook P, Ungar L, Trojanowski JQ, Grossman M. The power of neuroimaging biomarkers for screening frontotemporal dementia. Hum Brain Mapp 2014; 35: 4827–4840.

McMillan CT, Brun C, Siddiqui S, Churgin M, Libon D, Yushkevich P, et al. White matter imaging contributes to the multimodal diagnosis of frontotemporal lobar degeneration. Neurology 2012; 78: 1761–1768.

McNemar Q. Note on the sampling error of the difference between correlated proportions or percentages. Psychometrika 1947; 12: 153–157.

Medaglia JD, Lynall ME, Bassett DS. Cognitive network neuroscience. J Cogn Neurosci 2015; 27: 1471–1491.

Mori S, van Zijl PCM. Fiber tracking: principles and strategies - a technical review. NMR Biomed 2002; 15: 468–480.

Muldoon SF, Bridgeford EW, Bassett DS. Small-world propensity in weighted, real-world networks. arXiv preprint arXiv:150502194 2015.

Newman MEJ. Networks: An Introduction. Oxford University Press 2010.

Nir TM, Jahanshad N, Toga AW, Bernstein MA, Jack CR, Weiner MW, et al. Connectivity network measures predict volumetric atrophy in mild cognitive impairment. Neurobiol Aging 2015; 36: S113–S120.

National Institutes of Mental Health NI. Research Domain Criteria. http://www.nimh.nih.gov/research-priorities/rdoc/ 2015.

Onnela JP, Saramäki1 J, Kertész J, Kaski K. Intensity and coherence of motifs in weighted complex networks. Phys Rev E 2005; 71: 065103.

Raj A, Kuceyeski A, Weiner M. A network diffusion model of disease progression in dementia. Neuron 2012; 73: 1204–1215.

Rasmussen CE. Gaussian processes for machine learning. MIT Press 2006; .

Richiardi J, Achard S, Bunke H, Van De Ville D. Machine learning with brain graphs: Predictive modeling approaches for functional imaging in systems neuroscience. Signal Process 2013; 30: 58–70.

Rubinov M, Sporns O. Complex network measures of brain connectivity: uses and interpretations. NeuroImage 2010; 52: 1059–69. URL http://www.ncbi.nlm.nih.gov/pubmed/19819337.

Salvador R, Suckling J, Coleman MR, Pickard JD, Menon D, Bullmore E. Neurophysiological architecture of functional magnetic resonance images of human brain. Cereb Cortex 2005; 15: 1332–1342.

Schmidt R, de Reus MA, Scholtens LH, van den Berg LH, van den Heuvel MP. Simulating disease propagation across white matter connectome reveals anatomical substrate for neuropathology staging in amyotrophic lateral sclerosis. NeuroImage 2016; 124: 762–769.

Scholkopf B, Sung KK, Burges CJC, F G. Comparing support vector machines with gaussian kernels to radial basis function classifiers. IEEE T Signal Proces 1997; 45: 2758–2765.

Schroeter ML, Raczka K, Neumann J, Von Cramon DY. Neural networks in frontotemporal dementia?a meta-analysis. Neurobio Aging 2008; 29: 418–426.

Seeley WW, Crawford RK, Zhou J, Miller BL, Greicius MD. Neurodegenerative diseases target large-scale human brain networks. Neuron 2009; 62: 42–52.

Sporns O. Contributions and challenges for network models in cognitive neuroscience. Nat Neurosci 2014; 17: 652–660.

Stam CJ. Modern network science of neurological disorders. Nat Rev Neurosci 2014; 15: 683–695.

Suykens JAK, Vandewalle J. Least squares support vector machine classifiers. Neur Proc Letters 1999; 9: 293–300.

Toledo JB, Xie SX, Trojanowski JQ, Shaw LM. Longitudinal change in csf tau and aβ biomarkers for up to 48 months in adni. Acta Neuropathologica 2013; 126: 659–670.

Tustison NJ, Cook PA, Klein A, Song G, Das SR, Duda JT, et al. Large-scale evaluation of ants and freesurfer cortical thickness measurements. NeuroImage 2014; 99: 166–179.

van den Heuvel MP, Mandl RCW, Kahn RS, Hulshoff Pol HE. Functionally linked resting-state networks reflect the underlying structural connectivity architecture of the human brain. Hum Brain Mapp

2009; 30: 3127–41van den Heuvel MP, Sporns O. Network hubs in the human brain. Trends Cogn Sci 2013; 17: 683–696.

Wang H, Yushkevich PA. Multi-atlas segmentation with joint label fusion and corrective learning - an open source implementation. Front Neuroinfo2013; 7: 27.

Wang J, Wang L, Zang Y, Yang H, Tang H, Gong Q, et al. Parcellation-dependent small-world brain functional networks: A resting-state fMRI study. Hum Brain Mapp 2009; 30: 1511–1523.

Warren DE, Power JD, Bruss J, Denburg NL, Waldron EJ, Sun H, et al. Network measures predict neuropsychological outcome after brain injury. Proc Nat Acad Sci2014; 111: 14247–14252.

Wee CY, Yap PT, Zhang D, Wang L, Shen D. Constrained functional connectivity networks for mci classification. Med Image Comput Compute-Assist Intervent 2012; 7511: 212–219.

Whitwell JL, Avula R, Senjem MS, Kantarci K, Weingand SD, Samikoglu MS, et al. Gray and white matter water diffusion in the syndromic variants of frontotemporal dementia. Neurology 2010; 74: 1279–1287.

Zhou J, Gennatas ED, Kramer JH, Miller BL, Seeley WW. Predicting regional neurodegeneration from the healthy brain functional connectome. Neuron 2012; 73: 1216–1227.